\documentclass[%
 reprint,
 amsmath,amssymb,
 aps,
]{revtex4-1}

\usepackage{graphicx}% Include figure files
\usepackage{dcolumn}% Align table columns on decimal point
\usepackage{bm}% bold math

\begin{document}

\preprint{APS/123-QED}

\title{Size-polydisperse dust in molecular gas: Energy equipartition versus non-equipartition }

\author{ Alexander Osinsky, Anna S. Bodrova and Nikolai V. Brilliantov}
 \affiliation{Skolkovo Institute of Science and Technology, 121205, Moscow, Russia.}

\date{\today}

\begin{abstract}
We investigate numerically and analytically size-polydisperse granular mixtures immersed into a molecular gas. We show that the equipartition of granular temperatures of particles of different sizes is established; however, the granular temperatures significantly differ from the temperature of the molecular gas. This result is surprising since, generally, the energy equipartition is strongly violated in driven granular mixtures. Qualitatively, the obtained results do not depend on the collision model, being valid for a constant restitution coefficient $\varepsilon$, as well as for the $\varepsilon$ for viscoelastic particles. Our
findings may be important for astrophysical applications, such as protoplanetary disks, interstellar dust clouds, and
comets.
\end{abstract}

\maketitle

\section{Introduction}

What is common between very different objects, such as interstellar dust, protoplanetary discs \cite{protodust, prot1,
prot2}, comets \cite{comet} and dust devils on Mars \cite{devilMars}, Earth \cite{devilTheo} and possibly other
planets? All these systems are comprised of size and mass polydisperse dust particles immersed in molecular gas.
Interstellar molecular clouds possess high-density regions, so-called clumps. The density of matter there is large
enough to trigger a gravitational collapse, which eventually leads to the formation of stars. In the present study, we
assume that the density of matter is below this threshold.  Moreover, we assume that the granular gas is rarefied, so
that the surrounding molecular gas may be treated as a thermostat which is not affected by the granular gas.

The dust particles addressed here are macroscopic, but small enough grains, so that the gravitational interactions
between the grains may be neglected. Hence,  we have a granular mixture, driven by the molecular gas. Usually, in
granular mixtures, the energy equipartition between particles of different sizes is violated. This has been predicted
theoretically \cite{book,brey,dufty,GarzoDuftyMixture,Hrenya} and confirmed in experiments \cite{wildman,menon} and
computer simulations \cite{Hrenya,BRMotPRL,NatCom}. The same is true for such natural systems as Saturn rings, which
are essentially granular gas mixtures of particles with a size ranging from $10^{-3}$m to $1$m \cite{ringbook,pnas}.
The size polydispersity of the rings' particles stems from the permanent aggregation and fragmentation of the
constituents, which keeps the steady-state size distribution \cite{pnas,frank2004}. The energy equipartition for the
rings' particles does not hold as shown theoretically \cite{Ohtsuki1999,Ohtsuki2006} and in computer experiments
\cite{Salo1992b}.

Let a granular mixture contain $N$ different sorts of particles, of mass $m_k$ and diameter $\sigma_k$ with  $k=1,
\ldots N$. We assume that the particles are uniform spheres with the mass density of the material $\rho$, then $m_k
=\pi \sigma_k^3 \rho/6$. Without the loss of generality we assume that all masses are multiples of some minimal mass
$m_1$, that is, $m_k=km_1$. Let the number density of particles of mass $m_k$ be equal to $n_k$. We consider a space
uniform system, so that the particles of mass $m_k$ may be characterized by the velocity distribution function
$f\left(\textbf{v}_k, t\right)$. It quantifies the number of such particles with velocity $\textbf{v}_k$  at time $t$
in a unit volume. The number density may be expressed in terms of the distribution function as $n_k =\int
d\textbf{v}_kf\left(\textbf{v}_k, t\right)$. The average energy of particles of mass $m_k$ is characterized by the
corresponding granular temperature, $T_k$, defined as \cite{book,GarzoDuftyMixture},
\begin{equation}\label{grantemp}
\frac32 n_kT_k=\int d\textbf{v}_kf\left(\textbf{v}_k\right)\frac{m_kv_k^2}{2}.
\end{equation}
If all inter-particle collisions were elastic, the  energy equipartition between all sorts of the particles would hold.
The violation of equipartition stems from the dissipative nature of inter-particle collisions, which are quantified by
the restitution coefficient $\varepsilon$ \cite{book},
\begin{equation}
\label{rc} \varepsilon = \left|\frac{\left({\bf v}^{\,\prime}_{ki}
\cdot {\bf e}\right)}{\left({\bf v}_{ki} \cdot {\bf e}\right)}\right| \, ,
\end{equation}
where ${\bf v}^{\,\prime}_{ki}={\bf v}_{k}^{\,\prime}-{\bf v}_{i}^{\,\prime}$ and ${\bf v}_{ki}={\bf v}_{k}-{\bf
v}_{i}$ are the relative velocities of two particles after and before a collision, correspondingly, and ${\bf e}$ is a
unit vector connecting their centers at the collision instant. We do not consider very soft  particles where the
definition of the restitution coefficient is more subtle \cite{NegaEps}. The post-collision velocities are related to
the pre-collision velocities ${\bf v}_{k}$ and ${\bf v}_{i}$ as follows \cite{book}:
\begin{equation}\label{v1v2}
{\bf v}_{k/i}^{\,\prime} = {\bf v}_{k/i} + \frac{m_{\rm eff}}{m_{k/i}}\left(1+\varepsilon\right)({\bf v}_{ki} \cdot {\bf e}){\bf e} \, .
\end{equation}
Here $m_{\rm eff}=m_km_i/\left(m_k+m_i\right)$ is the effective mass of the colliding particles. To date, most studies
of granular gases have been focused on the case of a constant restitution coefficient
\cite{h83,gz93,dp03,dppre03,n03,nebo79,neb98,ap06,ap07}. This assumption contradicts, however, experimental
observations \cite{w60,bhd84,kk87}, along with basic mechanical laws \cite{rpbs99,titt91}, which indicate  that
$\varepsilon$ does depend on the impact velocity \cite{kk87,rpbs99,bshp96,mo97,sp98}. This dependence may be obtained
by solving the equations of motion for colliding particles with the explicit account for the dissipative forces acting
between the grains. The simplest first-principle model of inelastic collisions assumes viscoelastic properties of
particles' material, which results in viscoelastic inter-particle force \cite{bshp96} and finally in the restitution
coefficient \cite{rpbs99,sp98,delayed}:
\begin{equation}
\label{epsx}
\varepsilon_{ki} = 1 + {\sum^{20}_{j=1}}h_j\left(A\kappa_{ki}^{2/5}\right)^{j/2}\left |\left ({\textbf{v}}_{ki}\cdot\textbf{e}\right )\right |^{j/10},
\end{equation}
Here $h_k$ are numerical coefficients \cite{delayed}.
The elastic constant
\begin{equation}
\label{kappa}
\kappa_{ki} = \frac{\kappa}{\sqrt{2}}\frac{k+i}{k^{5/6}i^{5/6}\sqrt{k^{1/3}+i^{1/3}}}
\end{equation}
where
\begin{equation}
\kappa=\left (\frac{3}{2}\right )^{3/2}\frac{Y}{1-{\nu}^2}\left(\frac{6}{\pi\rho m_1^2}\right)^{\frac13}
\end{equation}
is a function of the Young's modulus $Y$ and Poisson ratio $\nu$; the constant $A$ quantifies the viscous properties of
the particles' material \cite{BPG_EPL,goldobin}:
\begin{equation}
\label{A}
A = \frac{1}{Y}\frac{\left(1+\nu\right)}{\left(1-\nu\right)}\left(\frac43\eta_1\left(1-\nu+\nu^2\right)+\eta_2\left(1-2\nu\right)^2\right)
\end{equation}
where ${\eta}_1$ and ${\eta}_2$ are the viscosity coefficients.

Recently we have shown that the distribution of granular temperatures in polydisperse mixtures of granular particles
follows the power law $T_k=T_1k^{\alpha}$, if the size distribution in a mixture is steep enough \cite{lev}. The
exponent $\alpha$ is universal for all steep size distributions for force-free granular mixtures. For driven granular
mixtures, $\alpha$ strongly depends on the agitation mode, in particular on the dependence of the driving force on the
particle size. In the current study, we investigate the distribution of temperatures in a mixture of granular particles
immersed into a molecular gas. We assume that the particles are very small, similar, but smaller, than the dust particles in sand
devils, tornado on Earth, interstellar dust, comets, and protoplanetary disks. In this case, the presence of a  molecular gas becomes important. Moreover, we assume that the action of the molecular gas on the granular mixture keeps it in a
steady state. 

The adhesion contact forces play an important role for small particles and they can aggregate at collisions, forming clusters. When such clusters collide at high impact speeds they can break into smaller pieces. If a steady state may be supported, these two processes  are balanced resulting in a steady  distribution of aggregates size \cite{pnas}. In the present study we limit ourselves to the range of parameters where aggregative (and respectively disruptive) collisions may be neglected.  The analysis of the conditions when the dust particles  undergo only bouncing collisions is given below.

We analyze both models of the restitution coefficient -- the simplified model of a constant
$\varepsilon$, as well as the realistic, first-principle model of visco-elastic particles. In each case, we obtain
qualitatively the same and somewhat unexpected result: The energy equipartition for the different granular species, along
with the strong violation of the equipartition between the granular mixture and molecular gas. It looks surprising since, generally, a strong violation of the energy equipartition in a driven granular mixture is expected \cite{lev}. Our
theoretical predictions have been checked by numerical simulations. Namely, we performed the Direct Simulation Monte
Carlo (DSMC) and confirmed the analytical findings. Interestingly, our conclusion supports a conjecture of the energy
equipartition in a granular mixture, immersed in molecular gas, proposed in Ref. \cite{spahn2006}.  The rest of the
study is organized as follows. In the next Section II, we specify the model and derive the granular temperatures for all
species, which is done for both models of the restitution coefficient. In Section III, we discuss the details of the
numerical simulations and compare the numerical and analytical results. Finally, in  Section IV, we discuss the application of our theory and summarize our findings.

%\begin{figure}\centerline{\includegraphics[width=0.8\columnwidth]{Gmix.png}}\caption{The granular particles of different masses $m_k$ (depicted with yellow circles) are  immersed into a molecular gas composed of molecules of mass $m_g\ll m_k$ (shown as small blue circles).} \label{Gmix}\end{figure}

\section{Temperature distribution in a granular mixture}
We consider a granular mixture, comprised on $N$ species of mass $m_k= k\, m_1$ immersed in a molecular gas with
temperature $T_g$ and molecular mass $m_g$. %(see Fig. \ref{Gmix}). 
Although being small, the dust particles are still much heavier than the gas molecules, that is, $m_g\ll m_1$. The collisions between the granular
particles and gas molecules are elastic. Since the velocity
distribution functions are close to Maxwellian distributions \cite{book}, we assume for simplicity that
$f_k(\textbf{v}_k,t)$ are Maxwellian:
\begin{equation}
f_k\left(\textbf{v}\right)=\frac{n_k}{\pi^{3/2}} \, \exp \left(- \frac{v_k^2}{v_{0,k}^2} \right),
\end{equation}
where $v_{0,k}=(2T_k/m_k)^{1/2}$ is the thermal velocity of particles of mass $m_k$. The distribution functions evolve according to the Boltzmann equation \cite{book},
\begin{equation}
\label{eq:BEgen} \frac{\partial}{\partial t} f_k\left(\textbf{v}_k, t \right) =  I^{\rm coll}_{k} + I^{\rm m.g.}_{k}.
\end{equation}
In Eq.~(\ref{eq:BEgen}) $I^{\rm coll}_{k}$ is the Boltzmann collision integral \cite{book}:
\begin{eqnarray}
\nonumber
I^{\rm coll}_{k} =\sum_{i=1}^{N}\sigma_{ki}^2\int d{\bf v}_{i} \int d{\bf e} \, \Theta (-{\bf v}_{ki}\cdot {\bf e}\,)\left|{\bf v}_{ki} \cdot {\bf e}\, \right|\,  \, \\
\left[\chi f_k({\bf v}_{k}^{\ \prime\prime},t)f_i ({\bf v}_{i}^{\ \prime\prime},t)-f_k({\bf v}_{k},t)f_i({\bf
v}_{i},t)\right] \label{II},
\end{eqnarray}
where $\sigma_{ki}=\left(\sigma_k+\sigma_i\right)/2$, with $\sigma_{k}=\left(6m_k/(\pi\rho)\right)^{1/3}$. The
summation is performed over all species in the system. ${\bf v}_{k}^{\, \prime\prime}$ and ${\bf v}_{i}^{\,
\prime\prime}$ are  pre-collision velocities in the so-called inverse collision, resulting in the post-collision
velocities ${\bf v}_{k}$ and ${\bf v}_{i}$.   The Heaviside step-function $\Theta(-{\bf v}_{ki}\cdot{\bf e})$ selects
the approaching particles and the factor $\chi$ equals the product of the Jacobian of the transformation $\left({\bf
v}_k^{\,\prime\prime}, \, {\bf v}_i^{\,\prime\prime}\right) \to \left({\bf v}_{k}, \, {\bf v}_{i}\right)$ and the ratio
of the lengths of the collision cylinders of the inverse and the direct collisions \cite{book}. In the case of a
constant restitution coefficient $\chi=1/\varepsilon^2$. For viscoelastic particles it has a more complicated form
\cite{book}; in what follows we do not need its explicit expression.

The second term $I^{\rm m.g.}_k$ describes the driving of the system due to collisions with the surrounding molecular
gas. It quantifies the energy injection into the granular mixture to compensate for its losses in dissipative collisions.
Since the mass ratio of the gas and grain particles $m_g/m_k$ is very small, the collision integral may be written
using the Kramers-Moyal expansion \cite{book}:
\begin{equation}\label{Iheat}
I^{\rm m.g.}_k=\frac{\partial}{\partial\bf v_k}\left(\gamma_k\textbf{v}_k+\bar{\gamma}_k\frac{\partial}{\partial
\textbf{v}_k}\right)f_k({\bf v_k},t).
\end{equation}

We investigate the evolution of granular temperatures, defined by Eq.~(\ref{grantemp}).  Multiplying the Boltzmann
equation (\ref{eq:BEgen}) by $m_kv_k^2/2$ for $k=1...N$ and performing integration over ${\bf v}_k$, we get the
following system of equations for evolution of the granular temperatures $T_k$ of species of different masses $m_k$:
\begin{eqnarray}
\left\{    \begin{array}{ll} \frac{d}{dt}T_1 = -T_1\sum_{i=1}^{N}\xi_{1i} +2\gamma_1\left(T_g-T_1\right) \cr \ldots\cr
\frac{d}{dt} T_k= -T_k\sum_{i=1}^{N}\xi_{ki} +2\gamma_k\left(T_g-T_k\right)\cr \ldots\cr \frac{d}{dt}T_N =
-T_N\sum_{i=1}^{N}\xi_{Ni} +2\gamma_N\left(T_g-T_N\right).
\end{array}\right.
\label{sys}
\end{eqnarray}
Here $\gamma_k=\gamma_{0k}\sqrt{T_g}$, $\gamma_{0k}=\frac{4}{3}n_g\sigma_k^2\sqrt{2\pi m_g}/m_k$ and
$\bar{\gamma}_k=\gamma_kT_g/m_k$ \cite{book,spahn2006}. The cooling rates $\xi_{ki}$ describe the decrease of
temperature of the species of mass $m_k$  due to collisions with the species of mass $m_i$. For the case of a constant
restitution coefficient these quantities read \cite{lev}:
\begin{eqnarray}\nonumber
\xi_{ki}(t) =\frac{8}{3}\sqrt{2\pi}n_i\sigma_{ki}^{2}\left(\frac{T_km_i+T_im_k}{m_im_k}\right)^{1/2} \left(1+
\varepsilon\right)\\ \left(\frac{m_i}{m_i+m_k}\right)\left[1-\frac{1}{2}\left(1+
\varepsilon\right)\frac{T_im_k+T_km_i}{T_k\left(m_i+m_k\right)} \right] . \label{xikconst}
\end{eqnarray}
We assume that the restitution coefficient $\varepsilon$ is the same for the collisions of particles of all sizes. In
the case of viscoelastic particles the cooling rates have the form:
\begin{eqnarray}\nonumber
\xi_{ki}(t) =\frac{16}{3}\sqrt{2\pi}n_i\sigma_{ki}^{2}\left(\frac{T_km_i+T_im_k}{m_im_k}\right)^{1/2}\!\left(\frac{m_i}{m_i+m_k}\right)\times\\\left[1-\frac{T_km_i+T_im_k}{T_k\left(m_i+m_k\right)}+\sum_{n=2}^{20}B_n\left(h_n-\frac12\frac{T_km_i+T_im_k}{T_k\left(m_i+m_k\right)}A_n\right) \right]\nonumber\\\label{xikvisc}\end{eqnarray}
where $A_n=4h_n+\sum_{j+k=n}h_jh_k$ are pure numbers and
\begin{eqnarray}
B_n(t)\!= \!\left(A\kappa_{ki}^{2/5}\right)^{\frac{n}{2}}\!
\left(\frac{2T_k}{m_k} +\frac{2T_i}{m_i} \right)^{\frac{n}{20}}\!\left(\frac{\left(20+n\right)n}{800}\right)
\!\Gamma\left(\frac{n}{20}\right)\nonumber\\\label{Bnk} \end{eqnarray}
with $\Gamma\left(x\right)$ being the Gamma-function. Driven
granular systems rapidly settle into a non-equilibrium steady state and all granular temperatures attain, after some time constant values, so that $dT_k/dt=0$. The system of equations (\ref{sys}) turns then into a set of algebraic
equations,
\begin{equation}\label{Tsteady}
T_k\sum_{i=1}^{N}\xi_{ki} =2\gamma_k\left(T_g- T_k\right)\,.
\end{equation}

Let us assume that the size distribution of the dust particles is steep enough and the number density of the granular mixture scales according to the power law,  $n_k=n_1k^{-\theta}$, with  $\theta >2$. Let the distribution of granular temperatures also scale according to the power-law: $T_k=T_1k^{\alpha}$. Then the following approximate relation holds \cite{lev}:
\begin{equation}
\label{xiint} \sum_{i=1}^{N}\xi_{ki}  \sim \left\{ \begin{array}{ll}         k^{\frac{\alpha}{2}- \frac56} \int_1^N
i\, n_i \, di & \mbox{if } \, \, \alpha \geq 1 \\         {} \\         k^{- \frac13} \int_1^N i^{\frac{\alpha +1}{2}}
\, n_i \, di & \mbox{if } \, \, 0< \alpha < 1 \, .
\end{array}\right.
\end{equation}
Substituting Eq.~(\ref{xiint}) into Eq.~(\ref{Tsteady}) and taking into account that $\gamma_k=\gamma_1k^{-1/3}$, we
conclude that the system \eqref{xiint} is compatible only for $\alpha=0$. This immediately implies the equality of the
granular temperatures,  $T_k=T_1$, that is, the energy equipartition, and justification of the conjecture of Ref.
\cite{spahn2006}.

We present the sum over the cooling rates in the form
\begin{equation}
\sum_{i=1}^{N}\xi_{ki}=\sqrt{T_1}\xi_{0k},
\end{equation}
where for a constant restitution coefficient:
\begin{eqnarray}\nonumber
&&\xi_{0k}=\sum_{i=1}^{N}\frac{8}{3}\sqrt{2\pi}n_i\sigma_{ki}^{2}\left(1+ \varepsilon\right) \sqrt{\frac{m_i}{m_k\left(m_i+m_k\right)}}\times\\
&&\qquad \left(1-\frac{1}{2}\left(1+ \varepsilon\right)\right)
\end{eqnarray}
and for viscoelastic particles:
\begin{eqnarray}\nonumber
&&\xi_{0k} =\sum_{i=1}^{N}\frac{16}{3}\sqrt{2\pi}n_i\sigma_{ki}^{2}\sqrt{\frac{m_i}{m_k\left(m_i+m_k\right)}}\times\\
&&\qquad \sum_{n=2}^{20}B_n\left(\frac12\sum_{j+k=n}h_jh_k-h_n\right)
\end{eqnarray}
where the coefficients $B_n(t)$ given by Eq.~(\ref{Bnk}) now take the form
\begin{eqnarray}
\nonumber &&B_n(t)\!=\left(A\kappa_{ki}^{2/5}\right)^{\frac{n}{2}}\left(2T_1\right)^{\frac{n}{20}}\left(\frac{m_i+m_k}{m_im_k}\right)^{\frac{n}{20}}\!\times\\
&&\qquad \left(\frac{\left(20+n\right)n}{800}\right)\!\Gamma\left(\frac{n}{20}\right).\nonumber
\end{eqnarray}
The granular temperature of the smallest particles (monomers) $T_1$ can be found from the equation:
\begin{equation}\label{EqT1}
T_1\sqrt{T_1}+b\sqrt{T_g}T_1-b \sqrt{T_g}T_g=0,
\end{equation}
where we introduce the notation, $b=2\gamma_{01}/\xi_{01}$. The quantity $b$ is a function of $T_1$ for granular
particles, colliding with velocity-dependent restitution coefficient, $b=b(T_1)$, so that Eq.~
(\eqref{EqT1}) is a transcendental equation. However  for the case of a constant restitution coefficient $b$ is constant. Introducing
$x=\sqrt{T_1/T_g}$, we recast Eq.~(\ref{EqT1}) in a cubic equation:
\begin{equation}
\label{x}
x^3+bx^2-b=0,
\end{equation}
with the solution in the form:
\begin{equation}
\label{x2}
x=\frac13\left(-b+zb^2+\frac{1}{z}\right)\,,
\end{equation}
where $z=2^{1/3} \left( 27 b -2 b^3+3\sqrt{3}\sqrt{27 b^2-4b^4} \right)^{-1/3}$. This yields the explicit expression
for temperatures of granular particles, colliding with constant restitution coefficient, in terms of the molecular gas temperature $T_g$:
\begin{eqnarray}
\label{Tempconst}
T_k&=& T_1=B\, T_g  \\ 
\label{B}
B&=&\frac{1}{9} \left(-b+zb^2+\frac{1}{z}\right)^2. \nonumber
\end{eqnarray}
It may be shown that $T_k=T_1$ is always smaller than $T_g$, which also follows from the physical nature of these
quantities.

For viscoelastic granular mixture $T_k=T_1$ may be found from the numerical solution of Eq. \eqref{EqT1}, where the
dependence of  $b$ on $T_1$ is to be taken into account. In both cases of constant $\varepsilon$, as well as for
$\varepsilon$ for viscoelastic particles, the energy equipartition for all granular species is observed. At the same
time, the granular temperature significantly differs from the temperature of the molecular gas,  $T_k > T_g$.
Physically this implies a steady energy flux from the molecular gas to the granular mixture, which permanently loses
energy in dissipative collisions.

\begin{figure*}\centerline{\includegraphics[width=0.99\columnwidth] {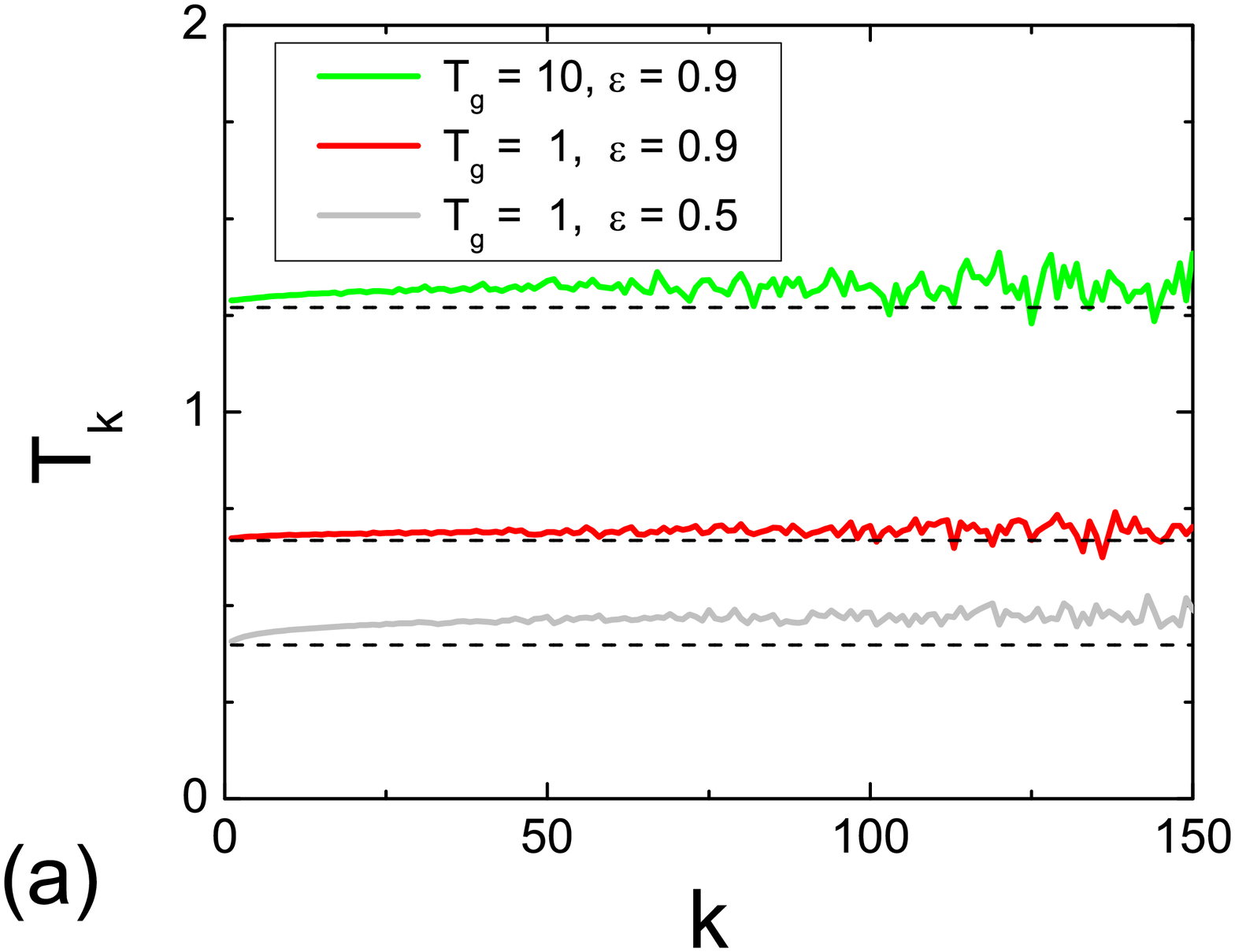}
\includegraphics[width=0.99\columnwidth]{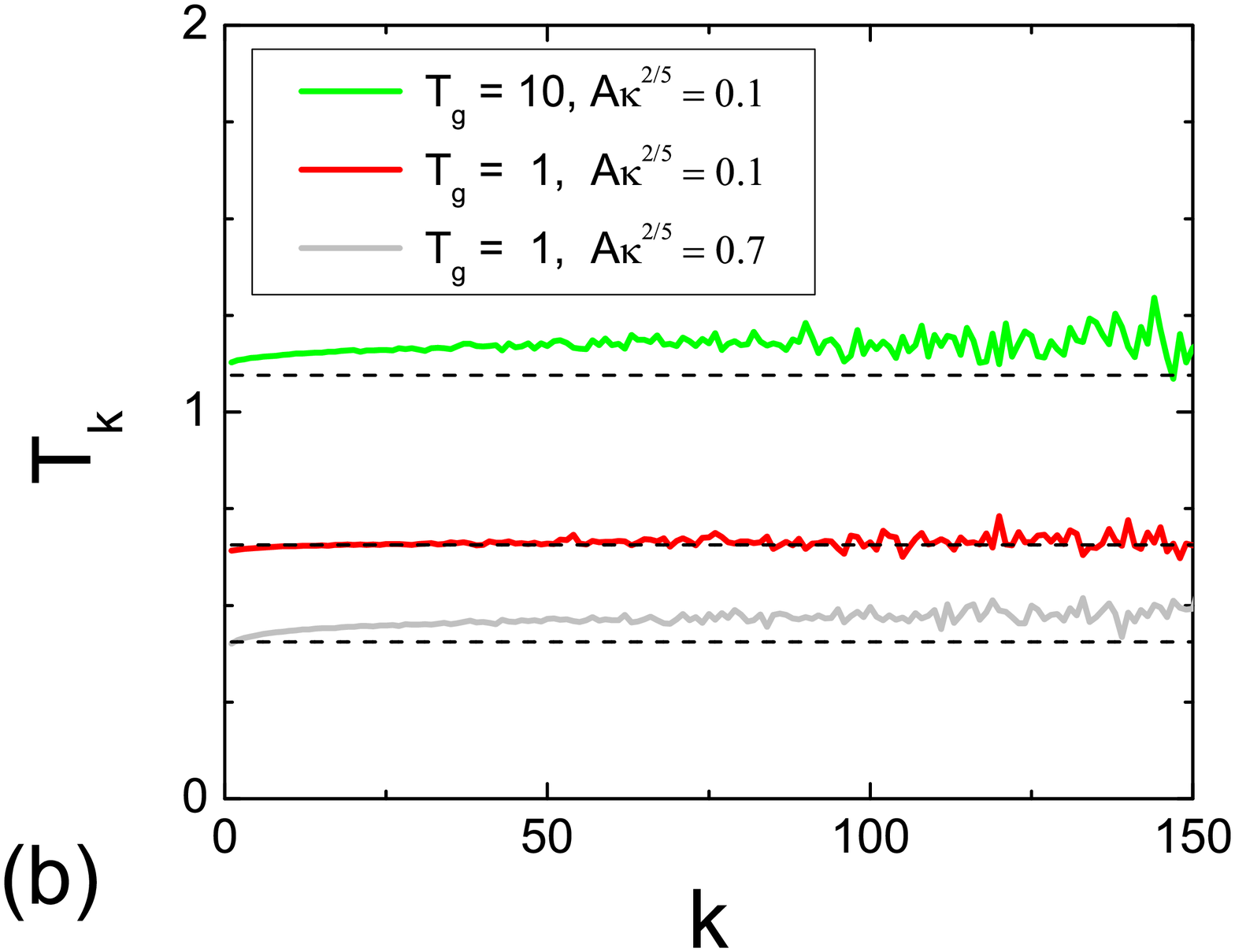}}
\caption{Granular temperatures $T_k$ of granular particles of mass $m_k$ immersed into a molecular gas of temperature
$T_g$. The granular particles collide with (a) constant restitution coefficient and (b) viscoelastic restitution
coefficient. The parameter values: $\sigma_1=1$, $m_1=1$ $\gamma_{01}=T_g^{-3/2}/2$, $n_k=n_1k^{-\theta}$ with $n_1=1$ and $\theta=3$. Colored
solid lines show simulation results, dashed black lines - analytical results for $T_1$ (Eq.~(\ref{Tempconst})) for
$\varepsilon=\rm const$ and numerical solution of Eq.~(\ref{EqT1}) for viscoelastic particles. One can clearly see that
the temperature distribution in granular mixture for both cases is close to the equipartition,  $T_k\simeq T_1$.
There is no energy equipartition between the granular mixture and molecular gas, $T_k<T_g$, since a permanent energy
flux from the gas to the mixture supports the steady state.
} \label{Gepsconst}
\end{figure*}

\section{Computer simulation and simulation results}

To check the prediction of our theory, we perform Direct Simulation Monte Carlo (DSMC), modified for the application to
multi-species systems. The detailed description of the DSMC  may be found elsewhere, see e.g. \cite{PoeschelBook}. Here
we briefly sketch some detail of the simulation method, with the focus on the  implementation of the thermostat, see
also \cite{Sant2000,WilliamsMacKintosh1996}.

We used
$N_1=3 \times 10^8$ particles of minimal mass $m_1$ (monomers), so that there were ${N_k} = \left\lfloor {{N_1}/{k^3}}
\right\rfloor = \left\lfloor {3 \cdot {{10}^8}/{k^3}} \right\rfloor $ particles of mass $m_k$ ($k$-mers).
Initially, the speeds of the particles are generated according to Maxwell distribution for $T=1$. Therefore, each particle $i$ is associated with its own mass $m_i$ and speed $\bf v_i$. Then the speeds change during the collisions and interaction with the molecular gas. The temperatures (determined by particles' kinetic energies) are measured after the system reaches a steady state. Numerically it can be determined by the time when the calculated temperatures stop changing monotonously.

The simulation of collisions between granular particles has been performed according to the following scheme:
\begin{enumerate}
\item Choose the sizes of colliding particles $i$ and $k$.
\item Choose the particles with speeds $\textbf{v}_i$ and $\textbf{v}_k$ with the probability,
proportional to $| (\textbf{v}_i - \textbf{v}_k) \cdot  \textbf{e} |$, where $\textbf{e}$ is the collision direction (a
random unit vector).
\item Update the speeds of the colliding particles according to the collision rules.
\end{enumerate}
While the particles' collisions are calculated one by one, interaction with molecular gas is performed simultaneously for all particles. We will describe it later in this section. For now, let us briefly go through each step.

1. The sizes of the colliding particles can be determined with the help of the upper bounds on the collision rates $C_{ik}$
\[
  C_{ik} = const \cdot N_i N_k \sigma_{ik} \left( |v_i|_{\max} + |v_k|_{\max} \right),
\]
where $\sigma_{ik} = \left( \frac{\sigma_i + \sigma_k}{2}\right)^2 \sim (i^{1/3}+k^{1/3})^2$ is the collision cross-section. Sizes $i$ and $k$ are selected with probability
\[
  p_{ik} = C_{ik}/{\mathop\sum\limits_{i,k=1}^{M} C_{ik}},
\]
where $M$ is the maximum particle mass in the system.

2. After the sizes are determined, two particles $j$ and $l$, corresponding to masses $i$ and $k$, are selected at random. Let us denote their speeds by $v_i^j$ and $v_k^l$. Then the collision is accepted if
\[
  | \textbf e (\textbf v_i^j - \textbf v_k^l)| > {\rm rand[0,1)} (|v_i|_{\max} + |v_k|_{\max}),
\]
where $\bf e$ is a random unit vector. Otherwise nothing happens and we choose the sizes again.

3. In case the collision is accepted, post-collision velocities are calculated, like in equation (\ref{v1v2}), with the appropriately defined restitution coefficient.

The action of the molecular gas is described by the term $I_k^{\rm m.g.}$ of the Boltzmann equation
\eqref{eq:BEgen}. Obviously, this term plays the role of a thermostat. In the lack of collisions between dust particles,
the equation for the distribution function reads,
\begin{equation}
  \frac{\partial f(\bf v_k, t)}{\partial t} = I_k^{\rm m.g.}.
\end{equation}
This equation, with $I_k^{\rm m.g.}$ given by Eq.~(\ref{Iheat}) is a Fokker-Planck equation, which corresponds to the
Langevin equation (see e.g. \cite{book}):
  \begin{eqnarray}
  &&\frac{d \textbf{v}_k}{dt} = - \gamma_k \textbf{v}_k + \textbf{F}_k^{st},  \label{eq:Lang}\\
  &&\left\langle \textbf{F}_k^{st} \right\rangle = 0, \quad
  \left\langle \textbf{F}_k^{st}(t) \textbf{F}_k^{st}(t') \right\rangle =
  \frac23 \hat{\rm I} {\bar \gamma}_k \delta (t - t') , \label{eq:Lang1}
  \end{eqnarray}
where $\hat{\rm I}$ is the unit matrix and a direct (dyadic) product is implied in the second part of Eq.
\eqref{eq:Lang1}. The solution of the stochastic Langevin equation may be written as
\begin{eqnarray}\nonumber
 && \textbf{v}_k(t + \Delta t) = \textbf{v}_k(t) e^{-\gamma_k \Delta t} +
 \xi \textbf{e} \cdot \sqrt{3 \frac{T_g}{m_k} \left( 1 - e^{-2\gamma_k \Delta t} \right)},\\
 &&\xi \sim {\cal N}(0,1), \label{vthch}
\end{eqnarray}
where ${\cal N}(0,1)$ denotes the normal distribution with zero mean and unit variance.

 Numerically the thermostat is implemented by changing the speeds according to
Eq.~(\ref{vthch}). Value of $\xi$ and random direction $\bf e$ are calculated independently for each particle of size $k$. The chosen time interval $\Delta t$ corresponds to the time of $Nh$ collisions, where $N$ is the
total number of particles. $h=0.1$ is a parameter, which should be sufficiently small, to guarantee that each particle
experiences several times the action of the thermostat  between collisions with other granular particles.

\begin{figure}\centerline{\includegraphics[width=0.99\columnwidth] {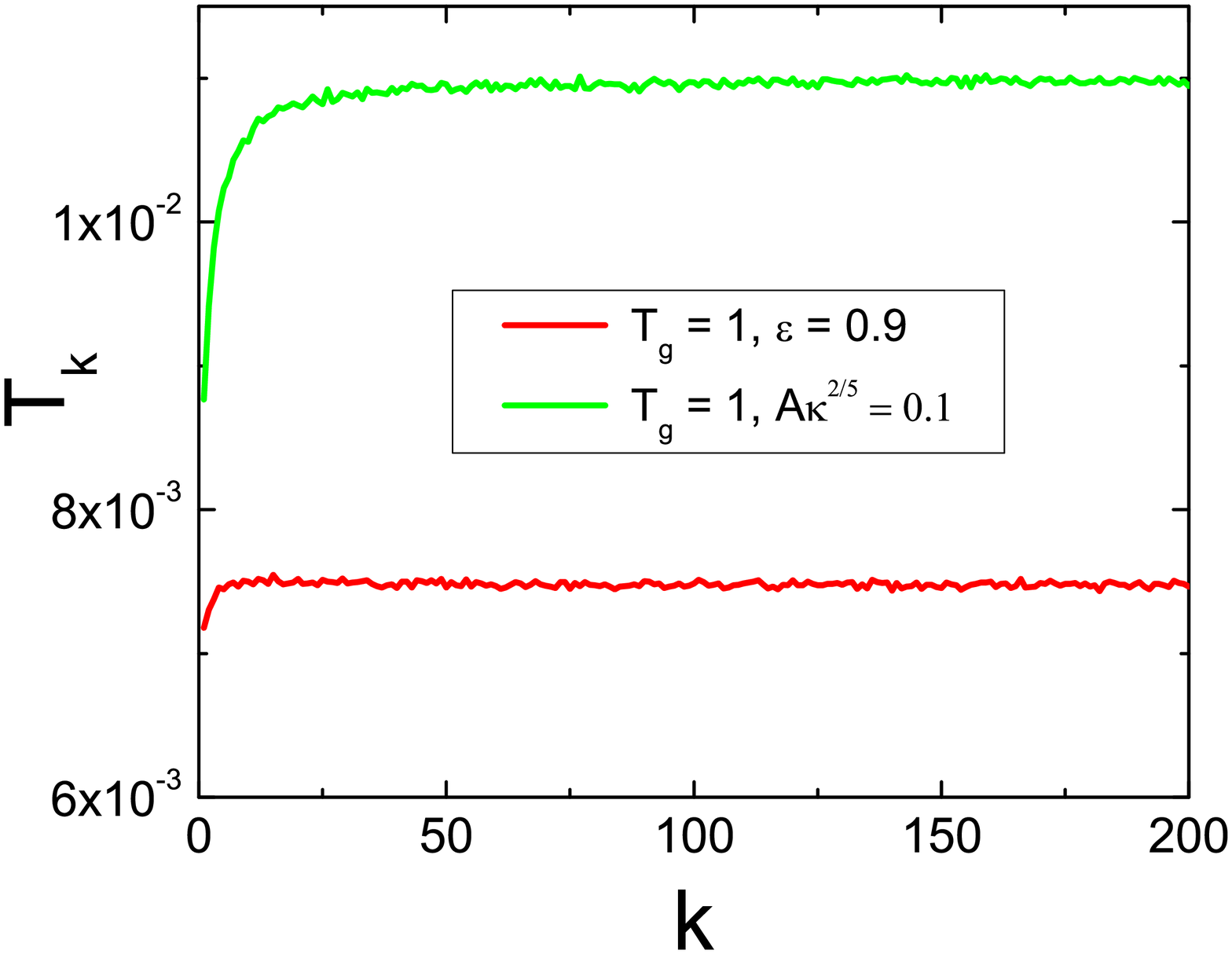}
}
\caption{The temperature distribution for the granular mixture with flat size distribution, $n_k = n_1 \Theta(k_{\rm max} -k)$, immersed into a molecular gas of temperature $T_g$. %Other parameter values are the same as at Fig.~\ref{Gepsconst}.
} \label{Gflat}
\end{figure}

The distribution of granular temperatures $T_k$, obtained by DSMC for different values of temperature of a molecular
gas $T_g$ and restitution coefficient $\varepsilon$  is given in  Fig.~\ref{Gepsconst}. Fig.~\ref{Gepsconst}a
corresponds to a constant restitution coefficient and in  Fig.~\ref{Gepsconst}b the distribution of granular
temperatures in a mixture of viscoelastic particles is shown for steep size distribution $n_k\sim k^{-\theta}$ with $\theta>2$. The granular temperatures rapidly tend to a steady-state
values, where the equipartition in a granular mixture is practically established: $T_k\approx T_1$ for any $k$. The
temperatures of the granular particles $T_k$ differ, however, from the temperature of the  molecular gas, $T_k<T_g$ as
it is predicted by the theory and expected from simple "physical" arguments of the heat flux from the molecular gas to
the granular mixture. The smaller the value of the restitution coefficient, the larger the difference between
temperatures of the  gas and the  mixture. For $\varepsilon=1$ both temperatures become equal and full equipartition is
established: $T_k=T_g$. Strictly speaking, the "true" equipartition should follow for $k \gg 1$, as Eq. \eqref{xiint}
is valid in this case. In practice, it is observed already for $k >10$. For other size distributions, e.g. for a flat
distribution, $n_k=n_1\Theta(k_{\rm max} -k)$, the DSMC also shows constant distribution of temperature, $T_k \approx T_1$ (see Fig.~\ref{Gflat}).

\section{Results and Discussion}
We investigate numerically  and theoretically a size-polydisperse granular gas mixture immersed into a molecular gas. We
assume that the molecular gas with temperature $T_g$ is not affected by the granular gas and plays the role of a
thermostat. The mixture  is comprised of $N$ different species of masses $m_k$ ($k=1, \ldots N$). We consider two models of dissipative collisions --
a simplified model of a constant restitution  coefficient, $\varepsilon ={\rm const.}$ and a realistic
model of viscoelastic particles, where the restitution coefficient depends on the relative velocities of colliding particles and their masses and sizes. For
both models, we observe qualitatively similar behavior:the granular mixture rapidly relaxes to
a steady state where granular temperatures of all species become equal, $T_k=T_1$ for all $k=1, \ldots N$, that is, the energy equipartition is observed. At the same time, the granular temperatures are not equal to the temperature of the molecular gas, $T_k< T_g$. This may be explained by the permanent energy flux from the molecular gas to the granular
mixture in the steady state, which compensates the energy losses in dissipative collisions of the grains. 

This effect resembles somehow the  experiments of $2D$ granular gas on a vertically vibrating substrate \cite{Swift_PRL,Aranson}. Here the difference between the "vertical" temperature $T_v$ and horizontal temperature $T_h$ is observed, $T_v>T_g$. The energy is injected into the vertical motion (analogy of the molecular gas) and is converted, though collisions, into the lateral motion (analogy of the dust motion in our system). Since the energy of the lateral motion is dissipated more intensively that the energy of the vertical motion, which is also pumped by the vibrations, the energy  equipartition breaks, that is $T_v>T_g$.

In our study we neglect the processes of collisional aggregation and fragmentation, assuming that only bouncing collisions take place. Let us estimate the range of parameters, where the conditions of purely bouncing collisions are fulfilled. We assume that sticking collisions occur  due to the adhesive interactions of the dust particles at a contact. The critical velocity, demarcating bouncing and sticking collisions has been reported in a number of studies, see e.g. \cite{frank2004,Brilliantov2007,Albers2006, dominik}. It is defined by the work against the adhesive forces $W_{ad}$, for which we use the explicit expression of Ref. \cite{Brilliantov2007}:
\begin{equation}
    \label{Wad} W_{ad}= q_0 (\pi^5\chi^5 R_{\rm eff}^4 D^2)^{1/3},
\end{equation}
where $q_0 = 0.09$ is a pure number, $\chi$ is the surface tension, $R_{\rm eff}= R_1R_2/(R_1+R_2)$ and $D=(3/2)(1-\nu^2)/Y$ (as previously, $\nu$ and $Y$ are the Poisson ratio and the Young  modulus). The condition of bouncing collisions reads, 
\begin{equation}
    \label{cond} T_1 =B\, k_BT_g > W_{ad},
\end{equation}
where $k_B$ is the Boltzmann constant (the temperature of the gas is in conventional units) and the quantity $B=B(b)$ is defined in Eq. \eqref{B}. From the definition of $b= 2\gamma_{01}/\xi_{01}$ and Eqs. \eqref{x} and \eqref{x2} follows that $B\sim 1$ for $b\sim 1$ and generally $B\sim b$. Moreover, $b\sim (n_g/n_1)\sqrt{m_g/m_1}$. 

Consider some typical quantities for a protoplanetary disk  \cite{AJ2006} (see also \cite{Kempf1999,AA2002,AJ2016}). It   contains the molecular gas $CO$ of molecular mass $40\, {\rm g/mole}$ and dust with size (radius) ranging from $5\cdot 10^{-3} \, \mu m$ to $10 \mu m$ \cite{AJ2006}.  We will use the following data for the material parameters of the dust particles \cite{Krijt2013,LougeEps,Kimura}: $Y=1 - 3\, GP$, $\rho=1 - 3 \cdot 10^3 \, {\rm kg/m^3}$, $\nu =0.25$ and   $\chi=0.002 - 0.025\, {\rm J/m^2}$, which corresponds to silica particles, including amorphous aggregates. For this parameters we obtain, that $B\sim b\sim 1$, provided the dust fraction belongs to the interval $n_1/n_g \sim 0.01\%  - 1\%$ for the dust particles of sub-micron
 to micron size, $\sigma \sim 0.01 - 1 \mu m$.  For the gas with temperature of $T_g =300\, K^o$, the dust particles of size $\sigma \sim 0.01 - 1 \, \mu m $ and smaller undergo pure bouncing collisions. At the same time, for the gas with temperature of $T_g=1500\, K^o$ the bouncing collisions experience particles of size $\sigma \sim 0.1 - 5 \, \mu m$ and smaller (the lower and upper limits here correspond to the according combination of the constants). 
 
 It is also interesting to estimate the relaxation time for the dust temperature, that is, the time needed to attain a steady-state temperature.  Taking into account that $T_g$ is commonly much larger than the dust temperature, Eq. \eqref{sys} for $T_1$ may be approximated as 
 \begin{equation}
    \dot{T}_1 \approx -2 \gamma_{1} (T_1-T_g),  
 \end{equation}
 which yields, $(T_1-T_g) \sim \exp(-t/\tau)$, with the relaxation time $\tau=2 \gamma_1$, where $\gamma_1$ has been defined after Eq. \eqref{sys}. Using again the data for a protoplanetary disc \cite{AJ2006}, $n_g \sim 10^{10} - 10^{16}\, m^{-3}$ and $T_g \sim 50 - 400 \, K^{o}$, we obtain that the relaxation time for particles of size $\sigma =1 \, \mu m$ ranges from $0.2$ hours  to $100$ years, while for particles of size $\sigma = 5\, \mu m$ from $1.5$ hour to $450$ years. 
 
 In conclusion, the observed energy equipartition in granular mixtures is surprising, since generally, the equipartition does not hold in driven granular gases with different particle sizes. The results of our study may be important to understand the properties of molecular gas-dust mixtures -- the systems, where small dust particles are immersed in the surrounding molecular gas. Especially our results may be useful to understand the properties of protoplanetary disks. 

\section{Acknowledgements} We gratefully acknowledge the usage of the Skoltech CDISE HPC cluster (Pardus).\\

%\newpage

\end{document}